\documentclass[11pt]{article}
\usepackage{amsmath,amssymb,color,epsfig,cite}
\usepackage{graphicx}
\usepackage{subfigure}
\usepackage{setspace}
\usepackage{amsfonts}
\newcommand{\hoch}[1]{$\, ^{#1}$}

\makeatletter
\@addtoreset{equation}{section}
\makeatother

\newcommand{\be}{\begin{equation}}
\newcommand{\ee}{\end{equation}}
\newcommand{\bea}{\setlength\arraycolsep{2pt} \begin{eqnarray}}
\newcommand{\eea}{\end{eqnarray}}

\def\fft#1#2{{\frac{#1}{#2}}}

\def\0{{\sst{(0)}}}
\def\1{{\sst{(1)}}}
\def\2{{\sst{(2)}}}
\def\3{{\sst{(3)}}}
\def\4{{\sst{(4)}}}
\def\5{{\sst{(5)}}}
\def\6{{\sst{(6)}}}
\def\7{{\sst{(7)}}}
\def\8{{\sst{(8)}}}
\def\sst#1{{\scriptscriptstyle #1}}

\begin{document}

\begin{center}
{\large {\bf Holographic Complexity Growth Rate in Horndeski Theory}}

\vspace{10pt}
Xing-Hui Feng \hoch{1} and  Hai-Shan Liu\hoch{2,1,3}

\vspace{10pt}

\hoch{1}{\it Center for Joint Quantum Studies, Tianjin University, Tianjin 300350, China}

\vspace{10pt}

\hoch{2}{\it Institute for Advanced Physics \& Mathematics,\\
Zhejiang University of Technology, Hangzhou 310023, China}

\vspace{10pt}
\hoch{3}{\it Center  for  Gravitation  and  Cosmology,  College  of  Physical  Science
and  Technology,  Yangzhou  University,  Yangzhou  225009,  China}

\vspace{10pt}

\vspace{30pt}

\underline{ABSTRACT}

\end{center}

Based on the context of complexity = action (CA) conjecture, we calculate the holographic complexity of AdS black holes with planar and spherical topologies in Horndeski theory. We find that the rate of change of holographic complexity for neutral AdS black holes saturates the Lloyd's bound. For charged black holes, we find that there exists only one horizon and thus the corresponding holographic complexity can't be expressed as the difference of some thermodynamical potential between two horizons as that of Reissner-Nordstrom AdS black hole in Einstein-Maxwell theory. However, the Lloyd's bound is not violated for charged AdS black hole in Horndeski theory.

\vfill {\footnotesize Emails: xhfeng@tju.edu.cn \, hsliu.zju@gmail.com }

\thispagestyle{empty}

\pagebreak

\tableofcontents
\addtocontents{toc}{\protect\setcounter{tocdepth}{2}}

%%%%%%%%%%%%%%%%%%%%%%%%%%%%%%%%%%%%%%%%

%\newpage
%%%%%%%%%%%%%%%%%%%%%%%%%%%%%%%%%%%%%%%%

\section{Introduction}
The gauge gravity duality states  connections between quantum gravity in string theory or more general settings and strongly coupled gauge field theories living on the boundary of the gravity background \cite{adscft1,adscft2,adscft3,adscft4}.  It has brought remarkable insight into understanding the phenomena various strongly coupled systems, such as low energy QCD, quark gluon plasma and condensed matter theory \cite{xxt1,xxt2,xxt3,xxt4}. Recently, two proposals about quantum computational complexity have emerged, namely, the conjecture of complexity = volume (CV) \cite{cv1,cv2} and the conjecture of complexity = action (CA)\cite{ca1,ca2}. Many studies has been done about these two conjectures, see \cite{camyers,cav1,cvt1,sub0,cav2,cav3,cav4,cav5,cav6,cav7,cav8,cav9,cav10,cav11,cav12} and references there in. In this paper, we shall follow the CA conjecture. The CA conjecture has been proposed by Brown $et al$\cite{ca1,ca2}, which states that the quantum complexity of ground state of CFT is given by the classical action evaluated on the "Wheeler-DeWitt patch" (WDW), and the WdW patch is the spacetime region enclosed by future and past light rays started from a bulk Cauchy slice, reflected at the boundaries and then ended in another bulk Cauchy slice, $e.g.$,  see Fig 1 in section 2.  The conjecture thus reads
\be
C_A = \fft{\cal I}{\pi} \,.
\ee

It was found that the late time action growth of various neutral black holes in the Einstein gravity theory is proportional to  the black hole mass $M$ \cite{ca1,ca2},
\be
\fft{d {\cal I}}{dt} = 2 M \,.
\ee
It suggests that the neutral black hole saturates the Lloyd's bound on the rate of the computation \cite{ldbd} . Later, the gravitational action growth for charged and/or rotated  AdS black holes in Einstein gravity  were studied \cite{twoh1}, the result turns out to be
\be
\fft{d {\cal I} }{dt} = \big( M - \Omega J - \mu Q \big)_+  -  \big( M - \Omega J - \mu Q \big)_- \,,
\ee
where the $\Omega \,, J$ are the angular velocity and angular momentum of the black hole, while the $\mu \,, Q$ are the electrical potential and charge of the black hole. The result was further generalized  and the gravitational action growth can be written as the difference of the generalized enthalpy between the two corresponding horizons \cite{ptwoh}
\be
\fft{d {\cal I} }{dt} = \big( F + TS \big)_+  -  \big( F + TS \big)_-  = {\cal H}_+ - {\cal H}_-\,,
\ee
 where $F$ is the free energy, $T$, $S$ are the temperature and entropy of the black hole, and ${\cal H} = F + TS$ is the generalized enthalpy. It was pointed out that the result still holds for higher derivative theories. Recently, it was explicitly showed that the result is true  for AdS black holes  in the $f(R)$ gravity, massive gravity theories \cite{frcg,masgr,fr,frmas} and  Lovelock gravity \cite{lovelock}.

However, it was pointed out that the action growth expression is different for charged black hole with a single horizon in the Einstein-Maxwell-Dilaton and Born-Infeld theories \cite{oneh}. It is thus worthwhile taking a further step to explore the pattern of the action growth of charged AdS black holes with only one horizon in higher derivative gravity theory. In this paper, we shall study the action growth of AdS black holes in  Horndeski gravity theory.

Horndeski theory is a kind of higher derivative scalar-tensor theory which has the similar property of Lovelock gravity, that the Largrangian involves terms which are  more than two derivatives, but the equations of motion are consisted of terms which have at most two derivatives acting on each field \cite{hd}. AdS black holes have been constructed in Horndeski gravities in \cite{hs1,hs2} and their thermodynamics were studied in \cite{th1,th2}. The stability and causality of these black holes were carried out in \cite{stab1,stab2,stab3}. Holographic application of Horndeski theory were investigated in \cite{holo1,holo2,holo3,holo4,hdap1,hdap2,hdap3,hdap4}, especially, it was shown in \cite{hdap3} that although there is no holographic $a$-theorem for general Horndeski gravity, there does exist a critical point in parameter space where the holographic $a$-theorem can be achieved, which suggests the Horndeski theory should have a holographic field theory dual. The AdS black holes we studied in this paper are in this critical point.

We shall study the action growth of unchanged black holes in $D=4$ Horndeksi gravity in section 2, and in section 3, we compute the action growth of  the charged black holes in the four dimensional Einstein-Maxwell-Horndeski gravity. We conclude our results in section 4.

\section{Neutral black holes in Horndeski theory}
In this section, we consider Einstein-Horndeski theory in four space time dimensions, which is given by
\be
{\cal I } = \fft{1}{16 \pi} \int dx^4 \sqrt g \Big[ \kappa (R - 2 \Lambda )   - \fft12 (\alpha g^{\mu\nu} - \gamma G^{\mu\nu}) \partial_\mu \chi \partial_\nu \chi     \Big]\,.
\ee
$G_{\mu\nu} = R_{\mu\nu} - \fft12 R g_{\mu\nu} $ is Einstein tensor and $\chi$ is a scalar field. For static ansatz
\be
ds^2 = - h(r) dt^2 + \fft{dr^2}{f(r)} + d\Omega^2_{2\,,\epsilon} \,, \quad \text{and} \quad \chi = \chi(r) \,,
\ee
the theory admits asymptotic AdS black holes with planar ($\epsilon = 0$) and spherical ($\epsilon = 1$) topology \cite{hs1}.

\subsection{Planar black hole }
The metric profile of the planar black hole solution is the same as that of Schwarzschild-AdS black hole, the solution is given by
\bea
h = f = g^2 r^2 -\frac{\mu}{r} \,, \qquad
\chi' = \sqrt {\fft \beta f } \,,
\eea
with constrains
\be
\Lambda = -\frac{3 g^2 (\beta  \gamma +2 \kappa )}{2 \kappa } \,, \quad \alpha = 3 \gamma  g^2 \,.
\ee
The solution has two integration constants, $\mu$ is related to the black hole mass and $\beta$ is related to scalar field $\chi$ which should be positive.  The AdS radius $l = 1/g$ is not determined by the cosmological constant $\Lambda$ but by the ratio of $\alpha$ over $\gamma$ which is precise in the critical point, and thus the holographic $a$-theorem holds for this system as pointed out in \cite{hdap3}.  When $\mu = 0$, the solution turns out to be an AdS vacuum with the scalar $\chi$ being logarithmic in terms of radial coordinate $r$. The conformal symmetry of the AdS is broken down to the Poincare symmetry plus the scaling symmetry because of the logarithmic scalar $\chi$, which means the dual field theory is scaling invariant. The Horndeski coupling $\gamma$ doesn't have a smooth zero limit and should not be treated as a perturbative  parameter. It was also showed that the kinetic term of the scalar perturbation $\delta \chi$ is non-negative as long as $\gamma$ is great than zero \cite{hdap4}.

 The mass  of the black hole is given by, for more detail about the thermodynamics of the black hole we refer to  \cite{th1},
\be
M = \fft{4 \kappa + \beta \gamma }{32 \pi} \, \mu \,.
\ee

Now, we follow the method in \cite{camyers} to calculate the late time action growth. In this method, the null coordinates were introduced
\be
du = dt + \fft {1} {\sqrt{hf}}  \,, \quad dv = dt - \fft {1} {\sqrt{hf}} \,,
\ee
and
\be
u = t + r^*(r) \,, \quad v = t - r^*(r) \,, \quad \text{with} \quad r^*(r) = \int \fft{1}{\sqrt{hf}} dr \,.
\ee
The the metric can be written as
\be
ds^2 = - h du^2 + 2 \sqrt{\fft h f} du dr + r^2 d\Sigma^2 \,, \quad \text{or} \quad
ds^2 = - h dv^2 - 2 \sqrt{\fft h f} dv dr + r^2 d\Sigma^2 \,.
\ee
For the choices of $(t,r)$, $(u,r)$ and $(v,r)$, we have
\be
\int \sqrt g d^4 x = \Omega \int \sqrt{\fft h f} r^2 dr dw \,,
\ee
where $w = {t\,,u\,,v}$ and $\Omega$ is the volume of the two dimensional transverse space.

\begin{figure}[htp]
\begin{center}
\includegraphics[width=200pt]{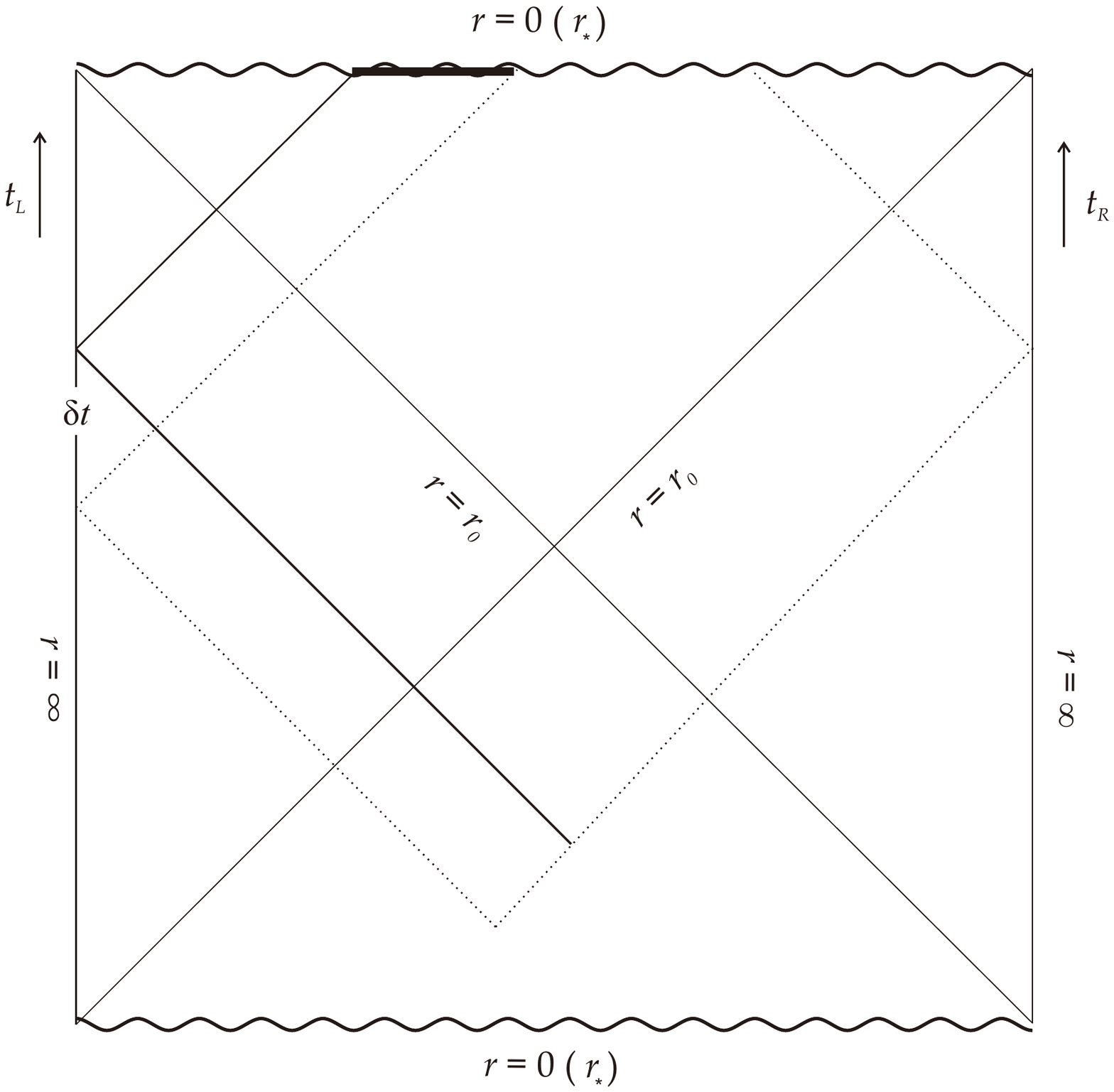}
\includegraphics[width=200pt]{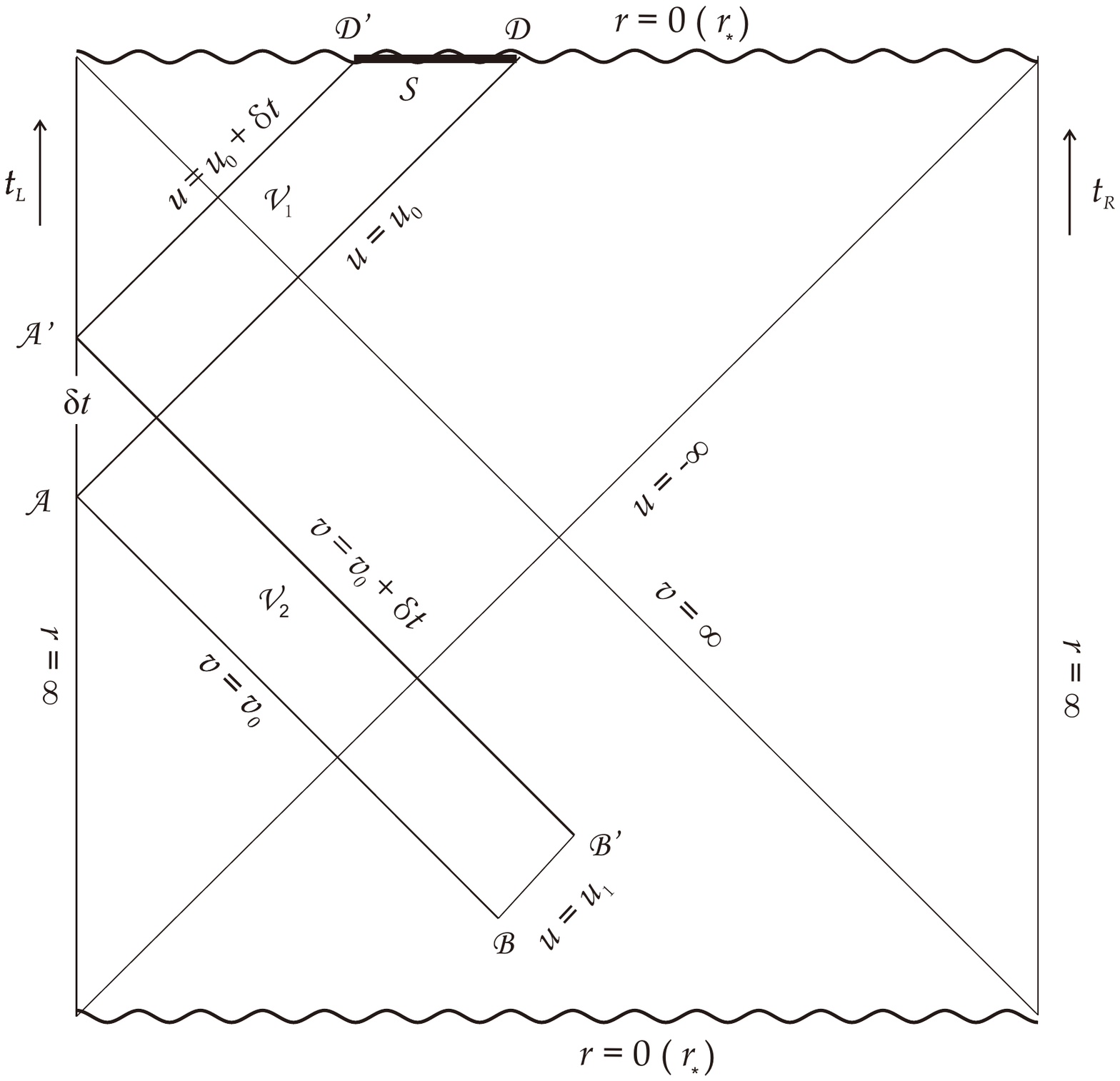}
\end{center}
\caption{\setstretch{1.0}\small\it Plots of WdW patches. The left panel shows the WdW patches at coordinate $t_0$ and $t_0+\delta t$, while the right panel shows the difference of the two patches. } \label{figure1}
\end{figure}

The Wheeler-de Witt patch of the black hole is defined by future light rays starting inside the black hole and reaching to the boundaries, then being joined to past light rays ending at the future singularity, which is illustrated in Fig.\ref{figure1}. The left part of Fig.\ref{figure1} shows the two patches corresponding to the actions ${\cal I}(t_0)$ and ${\cal I}(t_0 + \delta t)$ which have a time difference $\delta t$. In the right part of Fig.\ref{figure1}, the shadow area represents the difference of the actions $\delta {\cal I} = {\cal I}(t_0)- {\cal I}(t_0 + \delta t)$.The action difference $\delta {\cal I} = {\cal I}(t_0 + \delta t) - {\cal I}(t_0)$ is a sum of bulk, surface and joint contributions,
\be
\delta {\cal I} = {\cal I}_{bulk} + {\cal I}_{surf}+ {\cal I}_{joint} \,,
\ee
with
\be
{\cal I}_{bulk} = {\cal I}_{{\cal V}_1} - {\cal I}_{{\cal V}_2} \,, \quad {\cal I}_{surf} =  - \fft{ \kappa}{8 \pi}  \int _{r \rightarrow 0} K d\Sigma \,, \quad {\cal I}_{joint} =  \fft{ \kappa}{8 \pi} \oint _{B'} a  dS - \fft{ \kappa}{8 \pi} \oint _B a dS  \,.
\ee
Here, we choose the convention in \cite{camyers} that the contributions from null boundary are zero. $K$ is the Gibbons-Hawking term and $a$ will be illustrated later. Due to the equation of motion the Horndeski term in the action vanishes,  the bulk contribution has a simpler form
\be
{\cal I}_{{\cal V}_1} =   \fft{\beta \gamma - 2 \kappa}{16 \pi} \, g^2 \Omega \int_{u_0}^{u_0 + \delta t} \rho^3(u) du \,,
\ee
and
\be
{\cal I}_{{\cal V}_2} =  \fft{\beta \gamma - 2 \kappa}{16 \pi} \, g^2 \Omega \int_{v_0}^{v_0 + \delta t} \big[ \rho_0^3(v) - \rho_1^3(v) \big] dv \,,
\ee
where $\rho(u)$ is defined by $r^*(\rho) = \fft12 (v_0 + \delta t - u) $ and $\rho_{0,1}$ are defined by $r^*(\rho_{0,1}) = \fft 12 (v - u_{0,1})$.
By using the relation of the variables $u = u_0 + v_0 + \delta t - v$, the $\rho(u)$ term in ${\cal I}_{V_1}$ is the same as the $\rho_0(v)$ term in ${\cal I}_{V_2}$ and thus the two terms cancel out in the whole expression leaving
\be
{\cal I}_{V_1} - {\cal I}_{V_2} =  \fft{\beta \gamma - 2 \kappa}{16 \pi} \, g^2 \Omega \int_{v_0}^{v_0 + \delta t}  \rho_1^3(v)  dv \,.
\ee
Since it is a small variation from $v_0$ to $v_0+\delta t$, and so is the radius. Thus the above expression can be written as
\be
{\cal I}_{V_1} - {\cal I}_{V_2} =  \fft{\beta \gamma - 2 \kappa}{16 \pi} \, g^2 \Omega r_B^3 \delta t \,.
\ee
To the later time, the surface $B$ approaches black hole event horizon, it turns out to be
\be
{\cal I}_{V_1} - {\cal I}_{V_2} =  \fft{\beta \gamma - 2 \kappa}{16 \pi} \, g^2 \Omega r_0^3 \delta t\,,
\ee
where $r_0$ is the radius of the event horizon.

Next, we consider boundary terms involving $K$, the normal vector is $n_\alpha = \fft {1}{\sqrt f} \partial_\alpha r $, and $K$ is given by
\be
K = \nabla_\mu n^\mu = -\fft{1}{\sqrt{\fft h f}r^2} \partial_r( r^2 \sqrt{h} ) \,.
\ee
It is understood that the value in the square root should be absolute value, like $\sqrt h = \sqrt {|h|}$.
And
\be
- 2 \int_S K d\Sigma = 2  \kappa \Omega  \sqrt f \partial_r ( \sqrt h r^2 )\big|_{r \rightarrow 0} \delta t \,,
\ee
For our black hole solution, it is simply given by
\be
- \fft{ \kappa}{8\pi} \int_S K d\Sigma = \fft{3 \kappa}{16 \pi} \mu \Omega  \delta t \,.
\ee

Finally, we consider the joint contribution $\oint a dS $. $a$ is defined by
\be
a= ln( - \fft 12 k \cdot \bar k ), \quad \text{with}\quad k_\alpha = - c \, \partial_\alpha v \,, \quad \bar k_\alpha = \bar c \, \partial_\alpha u \,,
\ee
and $c,\bar c $ are positive constants. They are chosen to satisfy the asymptotic normalizations $k \cdot \hat t_L = - c $ and $\bar k \cdot \hat t_R = - \bar c $.  Then we have $ k \cdot \bar k  = 2 c \bar c / f$ and
\be
a = - ln ( \fft{- f}{c \bar c}  ) \,.
\ee
With those, we have
\be
\fft{ \kappa}{8 \pi} \oint _{B'} a  dS - \fft{ \kappa}{8 \pi} \oint _B a dS = \fft{\kappa}{16 \pi}  \Omega  \Big[ 2 r ln(\fft{-f}{c \bar c}) + \fft {f'}{f} r^2 \Big]\sqrt {hf} \big|_{r= r_B} \delta t \,,
\ee
At late times, $r_B \rightarrow r_0$, $r_0$ is the radius of the event horizon. The joint term is then given by
\be
 \fft{ \kappa}{8 \pi} \oint _{B'} a  dS - \fft{ \kappa}{8 \pi} \oint _B a dS = \fft{1}{16\pi}(2 \kappa g^2 r_0^3 + \kappa \mu ) \Omega \delta t \,.
\ee
Putting all the contribution together, we get the action difference
\be
\delta {\cal I}  = \fft{4 \kappa + \beta \gamma }{16 \pi} \mu \Omega \delta t = 2 M \delta t \,.
\ee
Thus, the action growth is
\be
\fft{\delta {\cal I}}{\delta t} = 2 M \,,
\ee
which is the same as that of Schwartzchild-AdS black hole in Einstein gravity. It is reasonable, since the profile of this neutral black hole in Horndeski theory, the boundary term and the on shell action are all the same as that of Einstein gravity. Next, we shall turn to the neutral black hole with spherical topology, the profile of which is totally different and more complicated.

\subsection{Spherical black hole }
The spherical black hole solution in Einstein-Horndeski theory was constructed in \cite{hs1}
\bea
\chi' &=& \sqrt { \fft{3 \beta g^2 r^2}{(3 g^2 r^2 + 1)} } \sqrt { \fft{1}{f}} \,, \quad
f = \frac{\left(3 g^2 r^2+1\right)^2 (\beta  \gamma +4 \kappa )^2}{\left(3 g^2 r^2 (\beta  \gamma +4 \kappa )+4 \kappa \right)^2} \, h   \,, \cr
h &=& g^2 r^2 + \frac{4 \kappa -\beta  \gamma }{\beta  \gamma +4 \kappa }-\frac{\mu }{r } + \frac{\left(\beta ^2 \gamma ^2\right) \arctan \left(\sqrt{3} g r\right)}{\sqrt{3} g r (\beta  \gamma +4 \kappa )^2}  \,.
\eea
The profile is more complicated and $h$ is not equal to $f$ any more. The mass of the black hole is
\be
M = \fft{4 \kappa + \beta \gamma }{32 \pi} \, \mu \,.
\ee
We shall follow the same method as that in the last subsection to calculate the action growth, the steps and situation are similar, hereafter we shall just present the main result of the calculation.
The WdW pataches are the same as the planar case, and the late time action difference is consist of three parts,
\be
\delta {\cal I} = {\cal I}_{bulk} + {\cal I}_{surf}+ {\cal I}_{joint} \,,
\ee
with
\be
{\cal I}_{bulk} = {\cal I}_{{\cal V}_1} - {\cal I}_{{\cal V}_2} \,, \quad {\cal I}_{surf} =  - \fft{ \kappa}{8 \pi}  \int _{r \rightarrow 0} K d\Sigma \,, \quad {\cal I}_{joint} =  \fft{ \kappa}{8 \pi} \oint _{B'} a  dS - \fft{ \kappa}{8 \pi} \oint _B a dS  \,.
\ee
The bulk contribution is
\be
{\cal I}_{V_1-V_2} = \frac{\mu  (\beta  \gamma +4 \kappa )^2-12 g^2 \kappa  r^3 (\beta  \gamma +4 \kappa )-16 \kappa ^2 r}{64 \pi (\beta  \gamma +4 \kappa )}\big|_{r \rightarrow 0}  \Omega \delta t\,.
\ee
The boundary part is
\be
- \fft{\kappa}{8 \pi} \int_{r=\epsilon}  K dS = \frac{3}{64 \pi} \mu  (\beta  \gamma +4 \kappa )  \Omega \delta t\,
\ee
and the joint contribution is
\be
 \fft{\kappa}{8 \pi} \oint_{B'}  a dS - \fft {\kappa}{8\pi} \oint_B a dS =\frac{\kappa  r \left(3 g^2 r^2 (\beta  \gamma +4 \kappa )+4 \kappa \right)}{16 \pi ( \beta  \gamma +4 \kappa ) } \big|_{r=r_0}  \Omega \delta t
\ee
With these, the total action difference is then given by
\be
\delta {\cal I} = \fft{\mu  (\beta  \gamma +4 \kappa ) \Omega}{16 \pi}\delta t = 2 M \delta t
\ee
So the action growth is
\be
\fft{\delta {\cal I}}{\delta t}  = 2 M \,,
\ee
which is the same as that of planar black hole, though the situation is quite different.

\section{Charged black holes in Horndeski theory}
In this section we consider Einstein-Hordeski-Maxwell theory in four dimensional space time. The theory is given by
\be
{\cal I } = \fft{1}{16 \pi} \sqrt g \Big[ \kappa (R - 2 \Lambda - \fft14 F^2 )   - \fft12 (\alpha g^{\mu\nu} - \gamma G^{\mu\nu}) \partial_\mu \chi \partial_\nu \chi     \Big]\,.
\ee
For static ansatz
\bea
ds^2 = - h(r) dt^2 + \fft{dr^2}{f(r)} + r^2 dx_i dx_i \,,
 \quad \chi = \chi(r) \,, \quad A = a(r) dt \,,
\eea
it was found that theory admits black hole solutions with planar and spherical topologies \cite{hs2}.

\subsection{Planar black hole}
First, we take a look at the charged planar black hole, the solution is given by
\bea
f &=& \frac{36 g^4 r^8 (\beta  \gamma +4 \kappa )^2}{\left(\kappa  q^2-6 g^2 r^4 (\beta  \gamma +4 \kappa )\right)^2}\, h \,, \cr
h &=& g^2 r^2 -\frac{\mu}{r} + \frac{\kappa  q^2}{r^2 (\beta  \gamma +4 \kappa )}-\frac{\kappa ^2 q^4}{60 g^2 r^6 (\beta  \gamma +4 \kappa )^2} \,, \cr
\chi' &=& \sqrt { \beta -\fft{\kappa  q^2}{6 \gamma g^2 r^4 } }\sqrt {\fft 1f } \,, \cr
a &=& a_0 -\frac{q}{r} +\frac{\kappa  q^3}{30 g^2 r^5 (\beta  \gamma +4 \kappa )} \,,
\eea
with constrains
\be
\Lambda = -\frac{3 g^2 (\beta  \gamma +2 \kappa )}{2 \kappa } \,, \quad \alpha = 3 \gamma  g^2 \,.
\ee
The thermodynamics was fully analysed in \cite{th1}. The mass, charge and electrical potential of the black hole are given by,
\be
M = \fft{4 \kappa + \beta \gamma }{32 \pi} \, \mu  \,, \quad T = \frac{6 g^2 r_0^4 (\beta  \gamma +4 \kappa )-\kappa  q^2}{8 \pi  r_0^3 (\beta  \gamma +4 \kappa )} \,, \quad \Phi_0 = a_0 \,, \quad Q = \fft{\kappa q}{16 \pi}\, ,
\ee
we chose the gauge that the electrical potential vanishes on the horizon.
Different from the neutral case, there is an additional curvature singularity $r_*$ where $f$ diverges
\be
\kappa  q^2-6 g^2 r_*^4 (\beta  \gamma +4 \kappa)  = 0 \,. \label{sing0}
\ee
 In order to describe a black hole, we require that the singularity $r*$ should be inside the event horizon, which insures that the temperature is always positive, $T>0$. From (\ref{sing0}) we can see that in the limit $q\rightarrow0$ the singularity $r_* \rightarrow 0$, going back to the usual singularity. However, it is worth pointing out that this charged black has no extreme limit, the solution has one and only one horizon. In order to see this property more clearly, we express the profile $h$ in terms of $r_*$,
\be
h = -\frac{3 g^2 r_*^8}{5 r^6}+\frac{6 g^2 r_*^4}{r^2}+g^2 r^2-\frac{\mu }{r} \,, \quad \text{and} \quad h' = \frac{18 g^2 r_*^8}{5 r^7}-\frac{12 g^2 r_*^4}{r^3}+2 g^2 r+\frac{\mu }{r^2}\,,
\ee
hereafter, a prime " ' "  denotes derivative with respect to " $r$ ".  We find that local extremes of profile $h$, where $h'(r_e)=0$, are equal to
\be
h_{e} = \frac{3 g^2 \left(r_e^4-r_*^4\right)^2}{r_e^6} \,,
\ee
which are greater than or equal to zero. When $r_e = r_*$, $h=0$, all the local extremes, singularity and event horizon degenerate. In this particular case, the solution describes a naked singularity rather than a black hole, which we shall not consider in this paper. So for a black hole solution, the extremes of $h$ are always positive. Since $h$ approaches $\infty$ as $r$ approaches $\infty$ and $h$ approaches $-\infty$ as $r$ approaches $0$, there should be at least one zero point. And as analysed before all the extremes are positive, thus there exists one and only one zero point for the profile $h$, which means that there is one and only one event horizon for the black hole solution, which is quite different from RN-AdS black hole.

We now turn to the calculation of action difference, the method we follow is the same as that of neutral case. The WdW patch is similar, too, except that the past light core end at the curvature singularity $r_*$ rather than the usual singularity $r=0$. Hence here, we shall skip the intermediate steps and present the final result. The total action difference is given by

%\be
%\delta S = S_{bulk} + S_{surf}+ S_{joint} \,.
%\ee

%The bulk contribution is
%\bea
%S_{bulk} = S_{V_1} - S_{V_2} =\big( F(r_0) - F(r^*) \big) \Omega \delta t\,,
%\eea
%where $F(r)$ is too complicated to show here.

%The boundary contribution
%\be
%S_{surf} = - \fft{\kappa}{8 \pi} \int_{r=r^*}  K dS \,,
%\ee
%is given by
%\be
%\frac{\kappa  \left(180 g^4 r^8 (\beta  \gamma +4 \kappa )^2+30 g^2 r^4 (\beta  \gamma +4 \kappa ) \left(2 %\kappa  q^2-3 \mu  r (\beta  \gamma +4 \kappa )\right)+\kappa ^2 q^4\right)}{80 \pi r (\beta  \gamma +4 %\kappa ) \left(6 g^2 r^4 (\beta  \gamma +4 \kappa )-\kappa  q^2\right)} \Big|_{r=r^*}  \Omega \delta t\,.
%\ee
%And the joints contribution is
%\be
%S_{joint} = \fft{\kappa}{8\pi} \oint_{B'}  a dS - \fft{\kappa}{8\pi} \oint_B a dS =\big( \frac{\kappa  %\left(6 g^2 r^4 (\beta  \gamma +4 \kappa )-\kappa  q^2\right)}{32 \pi r (\beta  \gamma +4 \kappa )} %\big)\big|_{r=r_0} \Omega \delta t \,.
%\ee
%We found that
%\be
%F(r_0) \Omega \delta t + S_{joint} = \fft{1}{16\pi} (\mu  (\beta  \gamma +\kappa ) - a_0 \kappa  q ) \Omega %\delta t \,,
%\ee
%and
%\be
%- \fft{\kappa}{8 \pi} \int_{r=r^*} \kappa K dS-F(r^*) \Omega \delta t =  - {\cal C}  \delta t + \fft{3 %\kappa \mu}{16 \pi}   \, \Omega \delta t\,,
%\ee
%where

%The total action difference is
\be
\delta {\cal I} = \fft{1}{16 \pi}\big(\mu  (\beta  \gamma +\kappa ) - a_0 \kappa  q \big )  \delta t -  {\cal C}_0  \delta t = \big( 2 M - \Phi_0 Q - {\cal C}_0 \big)  \delta t \,,
\ee
where
\be
{\cal C}_0 = \frac{g^2 r_*^3 (\beta  \gamma +4 \kappa )}{10 \pi } \,. \label{const0}
\ee
Thus the action growth is
\be
\fft{\delta {\cal I}}{\delta t} =  2 M - Q \Phi_0  - {\cal C}_0\,.
\ee
 As mentioned in previous, we can see from (\ref{sing0}) that  $r_*\rightarrow0$ when $q\rightarrow0$, thus, in the limit $q\rightarrow0$, the action growth $\fft{\delta S}{\delta t} \rightarrow 2 M$, going back to the neutral case as expected.  It is obvious from (\ref{const0}) that ${\cal C}_0$ is positive, so the action growth rate is less than $2 M$, satisfying the Lloyd's bound.

\subsection{Spherical black hole}
The solution is given by
\bea
f &=& \frac{4 r^4 \left(3 g^2 r^2+1\right)^2 (\beta  \gamma +4 \kappa )^2}{\left(\kappa  \left(q^2-8 r^2\right)-6 g^2 r^4 (\beta  \gamma +4 \kappa )\right)^2}\, h \,, \cr
h &=& g^2 r^2 + \frac{4 \kappa -\beta  \gamma }{\beta  \gamma +4 \kappa }-\frac{\mu }{r}+\frac{\kappa ^2 q^2 \left(3 g^2 q^2+16\right)}{4 r^2 (\beta  \gamma +4 \kappa )^2}-\frac{\kappa ^2 q^4}{12 r^4 (\beta  \gamma +4 \kappa )^2} \cr
&&+\frac{\arctan \left(\sqrt{3} g r\right) \left(2 \beta  \gamma -3 g^2 \kappa  q^2\right)^2}{4 \sqrt{3} g r (\beta  \gamma +4 \kappa )^2} \,, \cr
\chi' &=& \sqrt { \frac{6 \beta  \gamma  g^2 r^4-\kappa  q^2}{6 \gamma  g^2 r^4+2 \gamma  r^2} }\sqrt {\fft 1f } \,, \cr
a &=& a_0-\frac{\kappa  q \left(3 g^2 q^2+8\right)}{2 r ( \beta  \gamma +4 \kappa )}+\frac{\kappa  q^3}{6 r^3 ( \beta  \gamma + 4 \kappa )} \cr
&& -\frac{\sqrt{3} g q \arctan \left(\sqrt{3} g r\right) \left(3 g^2 \kappa  q^2-2 \beta  \gamma \right)}{2( \beta  \gamma +4 \kappa) } \,,
\eea
with constrains
\be
\Lambda = -\frac{3 g^2 (\beta  \gamma +2 \kappa )}{2 \kappa } \,, \quad \alpha = 3 \gamma  g^2 \,.
\ee
The mass, charge and electrical potential of the black hole are given by, for more detail about the thermodynamics of the black we refer to  \cite{th2},
\bea
M &=& \fft{(4 \kappa + \beta \gamma) \omega }{32 \pi} \, \mu\,, \quad  \Phi_0 = a_0 \,, \quad Q = \fft{\kappa q \omega }{16 \pi} \,, \cr
T &=& \frac{6 g^2 r_0^4 (\beta  \gamma +4 \kappa )-\kappa  \left(q^2-8 r_0^2\right)}{8 \pi  r_0^3 (\beta  \gamma +4 \kappa )} \,,
\eea
and $\omega$ is the volume of the unite sphere, we choose a gauge so that the electrical potential $a$ vanishes on the black hole event horizon.
Again, there is an additional singularity $r_*$ where $f$ diverges
\be
\kappa (q^2-8 r_*^2)-6 g^2 r_*^4 (\beta  \gamma +4 \kappa )  = 0 \,.
\ee
 In order to avoid  a naked singularity, we require that the singularity $r*$ should be inside the event horizon, which insures that the temperature is always positive, $T>0$.  It is similar to that of planar case, this charged black has no extreme limit,  too. The solution has one and only one horizon. With the same strategy, we can do the analysis by using $r_*$.  We find that local extremes of profile $h$, where $h'=0$, are equal to
\be
h_{e} = \frac{\left(r_e^2-r_*^2\right)^2 \left(3 g^2 \left(r_e^2+r_*^2\right) (\beta  \gamma +4 \kappa )+4 \kappa \right)^2}{r_e^4 \left(3 g^2 r_e^2+1\right) (\beta  \gamma +4 \kappa )^2} \,,
\ee
which are greater than or equal to zero. When $r_e = r_*$, $h=0$, all the local extremes, singularity and event horizon degenerate. In this particular case, the solution describes a naked singularity rather than a black hole, which we shall not consider in this paper. So for a black hole solution, the extremes of $h$ are always positive. Since $h$ approaches $\infty$ as $r$ approaches $\infty$ and $h$ approaches $-\infty$ as $r$ approaches $0$, there should be at least one zero point. And as analysed before all the extremes are positive, thus there exists one and only one zero point for the profile $h$, which means there is one and only one event horizon for the black hole solution.

Now we are in the position to calculate the action difference with the same procedure, the total action difference includes three parts, the bulk, boundary and joint parts and is given by

\be
\delta {\cal I} = \fft{1}{16 \pi} (\mu  (\beta  \gamma +\kappa ) - a_0 \kappa  q ) \omega \delta t -  {\cal C}_1 \delta t = \big( 2 M - Q \Phi_0  - {\cal C}_1   \big) \, \delta t \,.
\ee
where
\bea
{\cal C}_1 &=& \Big( -\frac{3 g^2 \kappa ^2 q^4}{64 \pi r (  \beta  \gamma +4  \kappa )}+\frac{g^2 r^3 (\beta  \gamma +4 \kappa )}{16 \pi }+\frac{\kappa ^2 q^4}{192 \pi  r^3 ( \beta  \gamma +  4   \kappa )}-\frac{r (\beta  \gamma -4 \kappa )}{16 \pi }  \cr
&&+ \frac{\sqrt{3} \arctan\left(\sqrt{3} g r\right) \left(4 \beta ^2 \gamma ^2-9 g^4 \kappa ^2 q^4\right)}{192 \pi  g (\beta  \gamma +4 \kappa )} \Big) \omega \Big|_{r=r_*} \,.
\eea
So the action growth is
\be
\fft{\delta {\cal I}}{\delta t}  = 2 M - Q \Phi_0  - {\cal C}_1 \,.
\ee
 Again, when $q\rightarrow 0$, $r_* \rightarrow \fft{q}{2 \sqrt 2}$  , thus in the limit of $q\rightarrow 0$, the combination $Q \Phi + {\cal C}_1$ approaches zero and  the action growth reduces to the neutral case, $\fft{\delta S}{\delta t} \rightarrow 2 M$. For small $q$ the combination of $Q \Phi + {\cal C}_1$ is given by
 \bea
 Q \Phi + {\cal C}_1 &\sim
  &\frac{\sqrt{2} q}{3 \pi  (\beta  \gamma +4)}  +\frac{q^2 \left(4-\sqrt{3} \beta  \gamma  r_0 \arctan \left(\sqrt{3} r_0\right)\right)}{16 \pi  r_0 (\beta  \gamma +4)} +\frac{q^3 (\beta  \gamma -4)}{32 \sqrt{2} \pi  (\beta  \gamma +4)} \cr
  && + \frac{q^4 \left(9 \sqrt{3} r_0^3 \arctan \left(\sqrt{3} r_0\right)+9 r_0^2-1\right)}{96 \pi  r_0^3 (\beta  \gamma +4)} + {\cal O}(q^5) \,.
 \eea
Here, we set $\kappa = 1$ and $g=1$ for simplicity. We can easily  see that the combination is great than zero for a not very small $r_0$. However, as we mentioned that the singularity should live inside the black hole event horizon, $r_0>r_*$, so $r_0$ can't be arbitrarily small, when $q$ is small, the black hole radus $r_0$ should be great than $\fft{q}{2 \sqrt 2}$, in this limit we have that
\be
 Q \Phi + {\cal C}_1 \sim \frac{2 \sqrt{2} q}{3 \pi  \beta  \gamma +12 \pi }-\frac{q^3 (\beta  \gamma -4)}{16 \left(\sqrt{2} \pi  (\beta  \gamma +4)\right)} + {\cal O}(q^5)\,,
\ee
which is obviously great than zero. So, we found that the combination of $Q \Phi + {\cal C}_1$ is positive for small $q$. For general parameter range we can not prove the combination $Q \Phi + {\cal C}_1$ is always greater than zero, however we plot $Q \Phi + {\cal C}_1$ as a function of $q$ and $r_0$ for large number of parameter choices which imply that $Q \Phi + {\cal C}_1$ is greater than zero, we present several of them in Fig. 2, . It seems that the combination of $Q \Phi + {\cal C}_1$ is always greater than zero, and the holographic complexity satisfies the Lloyd's bound.

\begin{figure}[htp]
\begin{center}
\includegraphics[width=200pt]{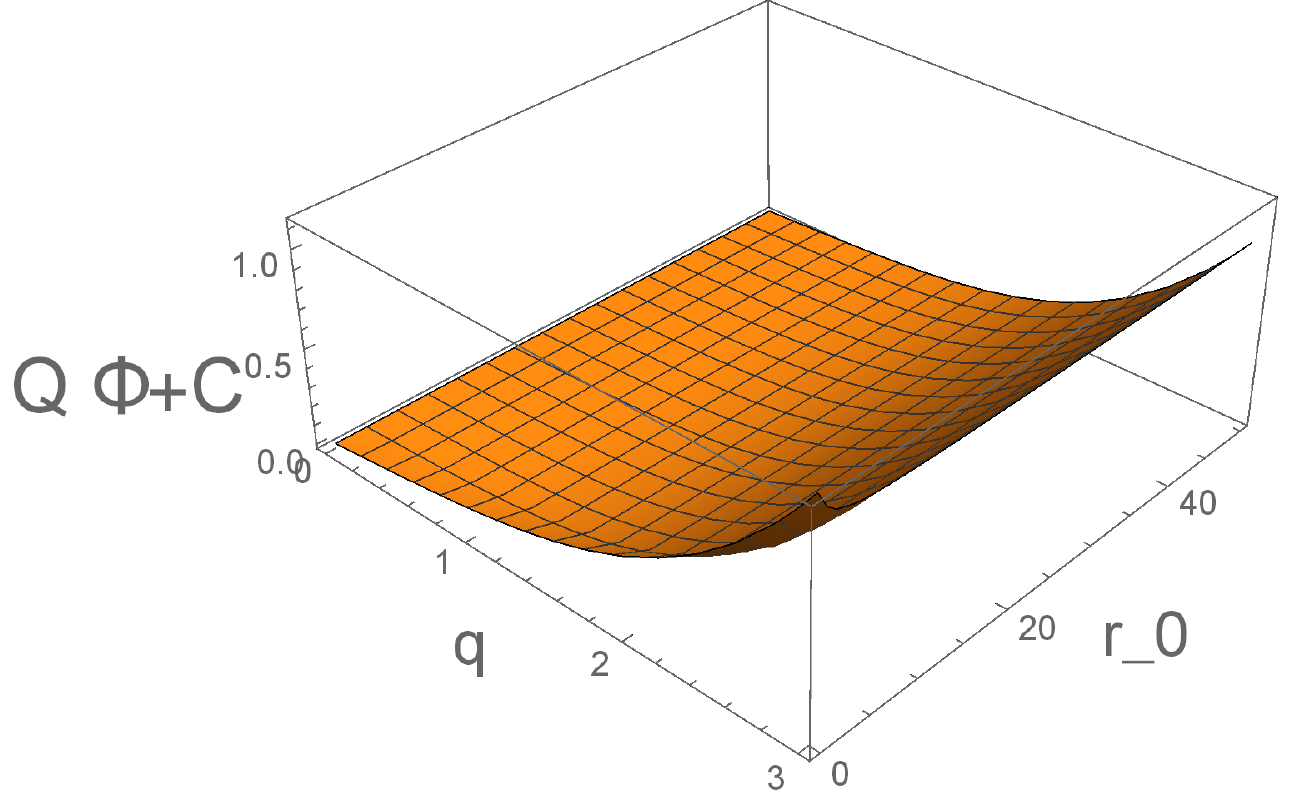}
\includegraphics[width=200pt]{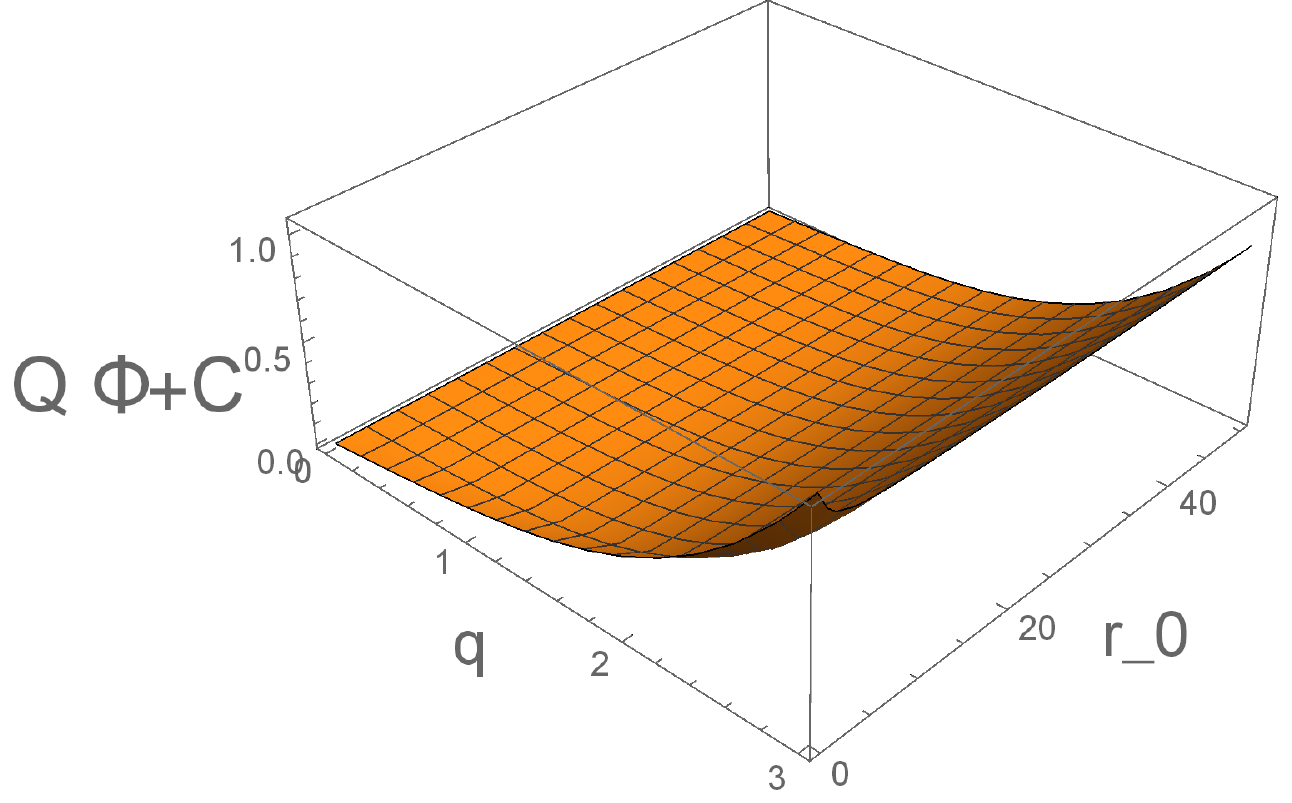}
\end{center}

\begin{center}
\includegraphics[width=200pt]{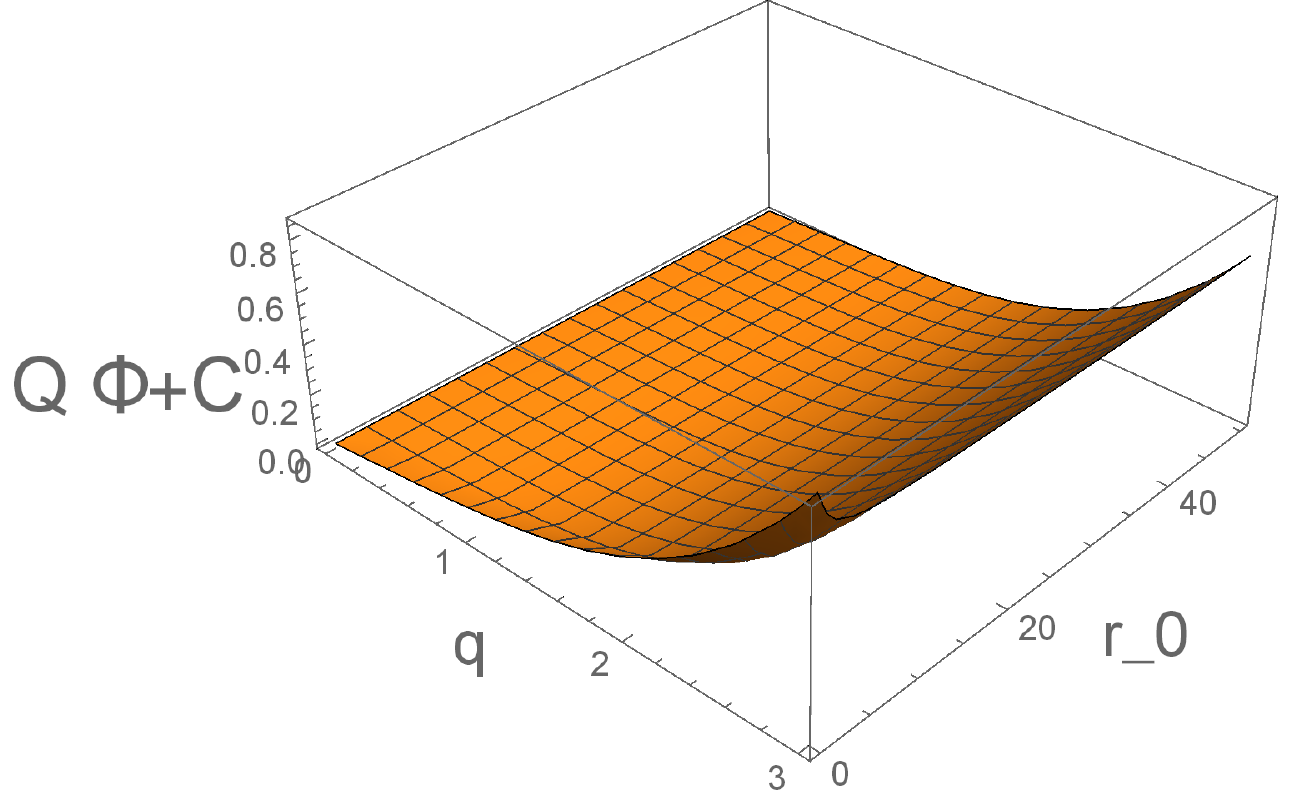}
\includegraphics[width=200pt]{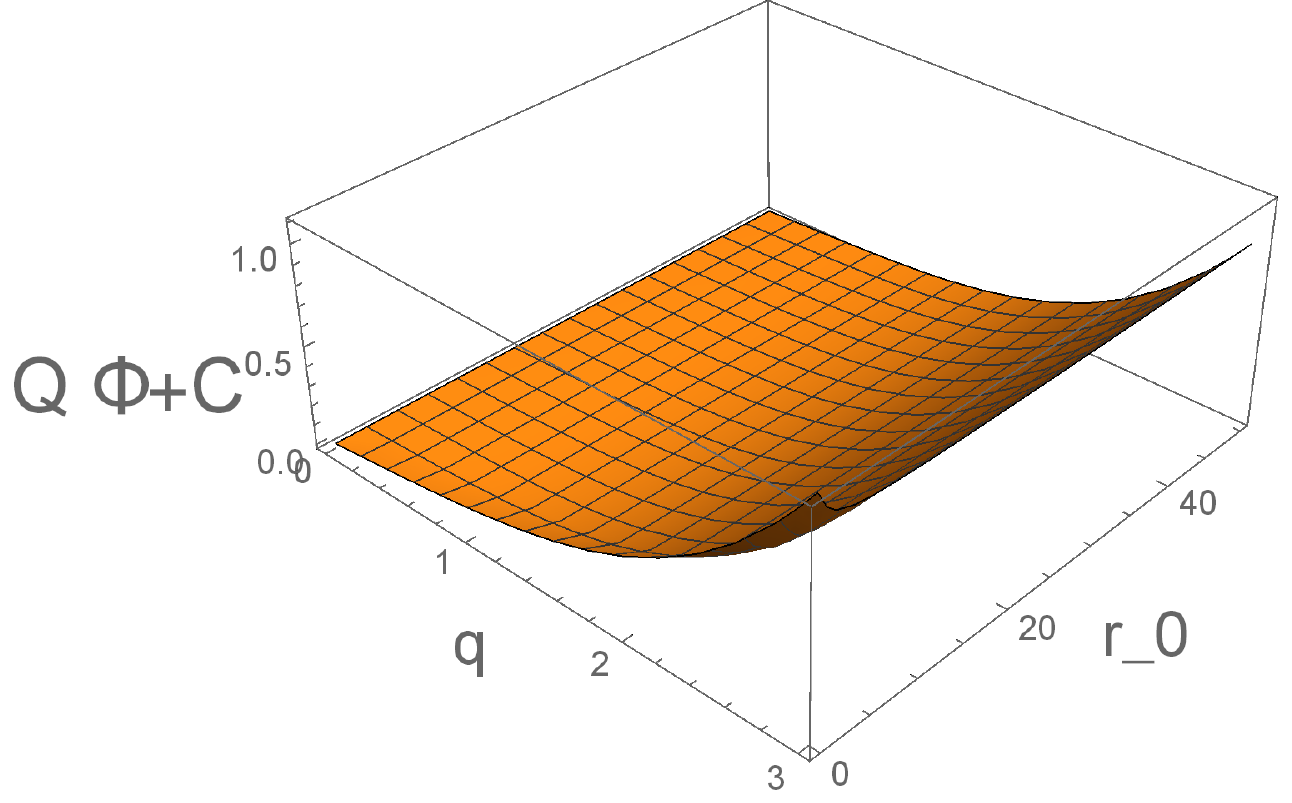}
\end{center}
\caption{\setstretch{1.0}\small\it Plots of $Q \Phi + {\cal C}_1$ for different $\beta\,, \gamma$ choices. Here we set $\kappa = 1$ and $g=1$ for all polts, while from top left panel to top right panel then down left panel  to down right panel, we set $(\beta\,, \gamma)$ equal to $(0.1\,,0.1)\,,(0.5\,,0.5)\,,(1\,,1)\, ,and \, (0.1\,,1)$ respectively. All panels show that $Q \Phi + {\cal C}_1$ is great than zero. } \label{figure2}
\end{figure}

\section{Conclusions}

In this paper, we studied the holographic complexity in Horndeski gravity theories through the " Complexity = Action" conjecture. In particular, we calculated the gravitational action growth of neutral and charged AdS black holes in Horndeski gravities. We found that the rate of change of action for neutral black holes with planar and spherical topologies is $2 M$ , which is  the same as the universal result\cite{ca1,ca2} and saturates the Lloyd's bound \cite{ldbd}.

 The charged black holes are more complicated. We analysed the metric profiles carefully and found that the charged black holes with planar and spherical topology both have only one event horizon which are quite different from that of the RN black holes. We computed the gravitational action growth for the charged black holes with planar and spherical topologies. It turns out that the action growth for planar topology is less then 2 $M$ thus satisfies the Lloyd's bound. Whilst, for spherical case, we showed that the action growth is less than 2$M$ when $q$ is small. Though we didn't prove analytically that the result holds for the whole range of parameters, we did numerically studies a substantial parameter choices and found that the action growth is less than $2 M$, which leads us to believe that the action growth is always less than $2M$ and satisfy the Lloyd's bound.

Here, we just studied the late time rate of change of holographic complexity, it is worthwhile going a step further to see the effect of the higher-derivative non-minimally coupled Horndeski term to the complexity of formation \cite{formation}, subregion complexity \cite{sub0,sub1, sub2,sub3} and also to explore full time dependence of the holographic complexity\cite{timedp1,timedp2,timedp3}.

\section*{Acknowledgement}

We are grateful to Luis Lehner and Hong Lu for useful discussion.  H.-S.L.~ is grateful for hospitality at Yangzhou University, during the course of this work.
X.-H.F.~ is supported in part by NSFC Grants No. 11475024 and No. 11875200. H.-S.L.~is supported in part by NSFC Grants
No.~11475148 and No.~11675144.

\end{document}